\documentclass[%
 aps,
 amsmath,amssymb,
 preprint,%
 superscriptaddress,%
]{revtex4-2}

\usepackage{graphicx}
\usepackage{dcolumn}
\usepackage{bm}
\usepackage{bm}
\DeclareMathAlphabet{\mathpzc}{OT1}{pzc}{m}{it}
\usepackage{tikz}

\begin{document}


\title{Stochastic evolution elasto-plastic modeling of a metallic glass}

\author{Bin Xu}
\affiliation{
  Department of Materials Science and Engineering, Mechanical Engineering, and Physics and Astronomy, Johns Hopkins University, Baltimore, Maryland 21218, USA
}
\author{Zhao Wu}
\affiliation{
  Institute of General Mechanics, RWTH Aachen University, Nordrhein-Westfalen, Eilfschornsteinstraße 18, 52062 Aachen, Germany
}%
\author{Jiayin Lu}
\affiliation{
  Department of Applied Mathematics, Harvard University, Cambridge, Massachusetts 02138, USA
}%
\author{Michael D. Shields}
\affiliation{
  Department of Materials Science and Engineering, Johns Hopkins University, Baltimore, Maryland 21218, USA
}%
\author{Chris H. Rycroft}
\affiliation{
  Department of Mathematics, University of Wisconsin, Madison, Wisconsin 53706, USA
}%
\author{Franz Bamer}
\affiliation{
  Institute of General Mechanics, RWTH Aachen University, Nordrhein-Westfalen, Eilfschornsteinstraße 18, 52062 Aachen, Germany
}
\author{Michael L. Falk}
\email{mfalk@jhu.edu}
\affiliation{
  Department of Materials Science and Engineering,  Mechanical Engineering, and Physics and Astronomy, Johns Hopkins University, Baltimore, Maryland 21218, USA
}%

\date{\today}

\begin{abstract}
This paper develops a general data-driven approach to stochastic elastoplastic modeling that leverages atomistic simulation data directly rather than by fitting parameters. The approach is developed in the context of metallic glasses, which present inherent complexities due to their disordered structure. By harvesting statistics from simulated metallic glass shear response histories, the material state is mapped onto a two-dimensional state space consisting of the shear stress and the inelastic contribution to the potential energy. The resulting elastoplastic model is intrinsically stochastic and represented as a non-deterministic dynamical map. The state space statistics provide insights into the deformation physics of metallic glasses, revealing that two state variables are sufficient to describe the main features of the elastoplastic response. In this two-dimensional state space, the gradually quenched metallic glass rejuvenates during the initial quasi-elastic shearing, ultimately reaching a steady-state that fluctuates about a fixed point in the state space as rejuvenation and aging balance.
\end{abstract}

\maketitle

Understanding and modeling plastic deformation of solids has been the focus of research for more than a century \cite{vonMises1913,Hill1948,Hill1979}.
In the context of single crystals, the bridge between the microscopic defect scale and the macroscopic behavior is relatively well understood, leading to highly developed theories of crystal plasticity \cite{Aravas1991,Anand1996}. Nonetheless, predictive plastic modeling remains a challenge in materials that exhibit substantial amounts of structural disorder. These include heavily dislocated crystals and polycrystals with small, potentially nanometer-scale, grains. Amorphous solids exist at the ultimate limit of disorder. The lack of any long-range order in these materials has stymied the formulation of widely applicable constitutive theories with high predictive capacity. Addressing this deficiency would provide tools to guide material and part design, as changes in composition and processing could be fundamentally related to mechanical response through a validated multiscale modeling procedure. Here, we formulate a new approach to plasticity modeling in such materials that is generalizable and can serve as a starting point for developing new methodologies.

The macroscale mechanism of the inelastic response of amorphous solids is driven by very localized elementary events that include a few hundred atoms while the surrounding material responds elastically.
Although such events collectively accommodate the strain of the macroscopic deformation exerted, they significantly vary from each other due to atomic-scale heterogeneity.
Consequently, a physically meaningful large-scale description of disordered solids needs to incorporate statistical information regarding sites of incipient plasticity, whose position and structure is \textit{a priori} unknown. 
One basic concept that has been deployed in a number of multiscale approaches is to define statistically representative volume elements (RVEs) \cite{mcdowell_perspective_2010,bargmann_generation_2018}, which are carefully chosen units prototypical of the essential physics of the material. An RVE is large enough if the results of increasing the cell size are predictable from the structural response of the RVEs at the measured size; determination of this minimal necessary size can be cumbersome \cite{kanit_determination_2003,chatterjee_prediction_2018}. Generally, RVEs may be defined from statistical data from simulations, experiments, or combinations of the two \cite{diehl_identifying_2017}. Having defined an RVE, one aims at modeling the response of larger-scale structures by coarse-graining these units so that the salient properties of the underlying physics are conserved or by incorporating RVEs into finite element schemes on the Gauss integration point level \cite{han_microstructure-based_2020}. Importantly, material response at scales above the RVE scale is treated deterministically. That is, the stochastic nature of the underlying deformation mechanisms is not accounted for.

This paper uses metallic glass (MG) as an exemplar for constructing a stochastic model of a disordered solid. MGs are a reasonable starting point because they can be relatively simple in their atomic-scale composition and bonding while they feature all the complexities of structurally disordered solids. Due to their excellent physical and mechanical properties they have received tremendous attention in the literature~\cite{Schuh2007MechanicalAlloys,Trexler2010MechanicalGlasses,Wang2004BulkGlasses,Hufnagel2016DeformationExperiments}, revealing particularly high strength~\cite{Inoue2003Cobalt-basedProperties, Wang2011Co-basedPlasticity}, high toughness~\cite{Demetriou2011AGlass, Gu2010CompressiveGlass, He2012Crack-resistanceToughness}, and, in some notable instances, high fatigue endurance~\cite{Gludovatz2013EnhancedMechanism,El-Shabasy2010FatigueGlass}. They have also shown some promise as materials for additive manufacturing applications \cite{wu_additive_2022}. While MGs can outperform classical crystalline metals in many regards, they often fail in a quasi-brittle manner due to nano-scale strain localization and the lack of longer length scale microstructural features able to redirect and distribute the strain. Although several phenomenological models can predict aspects of MG behavior~\cite{Lemaitre2002RearrangementsMaterials, Langer2004DynamicsTemperature, Spaepen1977AGlasses,Falk2004ThermalData,Anand2005AGlasses,Su2006PlaneSimulation,Anand2007ATemperatures,Demetriou2009Coarse-grainedMetals,Henann2008AForming,Johnson2002DeformationModel}, they are not able to capture the basic mechanisms of strain localization and failure observed in experiments. In particular, quantitatively predicting the microstructural evolution during plastic flow remains a challenge. To address this, significant research activities have focused on macro-scale engineering testing, mesoscale theory development, and simulations conducted at the atomic scale, but the critical linkages between these scales remain tentative~\cite{Hufnagel2016DeformationExperiments}.

Atomistic methods, most notably molecular dynamics (MD) and athermal quasi-static shear (AQS) simulations, have been used extensively to investigate MG mechanical response. While a complete review is not possible here, we note that early investigations characterized local structural fluctuations in MG~\cite{Srolovitz1983AnMetals,Egami1982LocalTransition,Deng1989SimulationIV}. Based upon further work, Falk and Langer motivated mechanical constitutive equations from the hypothesis that a point defect, the shear transformation zone (STZ), controls MG deformation~\cite{Falk2011DeformationMaterials, Falk1998DynamicsSolids}. Subsequent investigations studied localization and failure modes and the dependence of mechanical response on thermal history~\cite{Shi2007EvaluationSolids, Shi2005StructuralFilm, Shi2006DoesFailure, Shi2007Stress-inducedGlass, Shi2007SimulationsFilm,Bailey2004AtomisticProperties,Bailey2004SimulationStructure,Bailey2006AtomisticGlasses} as well as composition~\cite{Cheng2008AlloyingLiquids,Cheng2008LocalHistory,Cheng2008RelationshipAlloys,Cheng2009AtomicGlass}. More recent advances have investigated glass response in terms of energy landscapes~\cite{Cao2019PotentialRheology,Fan2017EnergyMaterial} and the physical underpinnings of mechanical transitions~\cite{Ozawa2018RandomMaterials} in ways that take advantage of new methods for equilibrating glass structures \textit{in silico}~\cite{Ninarello2017ModelsStudies}.

Much atomistic simulation work has focused on correlating aspects of atomic-scale properties with deformation. Detailed structural investigations revealed that certain topological bonding units correlate with deformation~\cite{Sheng2007PolyamorphismGlass}. Other analyses were able to correlate deformation with the most mobile atoms participating in eigenmodes associated with low-frequency vibrations in the phonon density of states~\cite{Derlet2012TheStructure,Schober1996Low-frequencyGlass, Rottler2014PredictingGlasses, Schoenholz2014UnderstandingDynamics}, which were shown to correlate with topological bonding measures~\cite{Ding2014SoftGlass}. Machine-learned short-range structural measures have been used to identify ``soft spots'' correlated with deformation events in both simulated and experimental systems~\cite{Cubuk2015IdentifyingMethods,Schoenholz2014UnderstandingDynamics,Schoenholz2017RelationshipSystems}. Thermal vibrations and non-linearities in atomic-scale response have also shown some predictive capacity for anticipating sites of plastic rearrangement~\cite{Gartner2016NonlinearSolids,Lerner2016MicromechanicsModes,Zylberg2017LocalGlasses}. Recent studies have further revealed that certain topological defects in the low frequency eigenmodes may be directly related to STZs~\cite{wu2023topology,desmarchelier2024topological}.

The question of how best to formulate a constitutive theory of MG deformation response motivates many of these investigations. In the past, we have pursued the construction of such models in the form of continuous and deterministic partial differential equations, most notably the effective temperature STZ theory~\cite{Falk1998DynamicsSolids,Falk2011DeformationMaterials,Bouchbinder2009NonequilibriumDeformation,Bouchbinder2009NonequilibriumTheory,Bouchbinder2009NonequilibriumPlasticity,Hinkle2017CoarseSolids,Kontolati2021}. Hinkle et al.~\cite{Hinkle2017CoarseSolids} and Kontolati et al.~\cite{Kontolati2021} previously devised methods to relate such theories to atomistic studies. We have found that, in many instances, such continuous, differentiable representations are not able to provide a satisfying mathematical description of the inherently discontinuous stochastic dynamics of the MG.
Here we instead draw upon a wide range of literature on elasto-plastic modeling~\cite{Bulatov1994ALocalization,Hebraud1998Mode-CouplingMaterials,Homer2009MesoscaleDynamics,Baret2002ExtremalPlasticity,Nicolas2014RheologyTemperature}, recently reviewed by Nicolas et al.~\cite{Nicolas2018DeformationModels}, to motivate an approach we will refer to as a stochastic evolution elastoplastic model (SEEM). In doing so, we start from the empirical observation common to many such models that deformation in amorphous materials arises from intermittent stress drops.
Our approach is closely related to two recent advances, the StEP model~\cite{Zhang2022Structuro-elasto-plasticitySolids}, an elasto-plastic model based on the evolution of ``softness'' obtained via machine learning, and a recent effort to draw close comparisons between one particular elastoplastic model and atomistic simulations of amorphous solids subjected to shear by comparing the evolution of their glassy landscapes~\cite{Kumar2022MappingModel}. The most salient distinguishing feature of our approach is that, rather than proposing an elastoplastic formulation based on a physically motivated \textit{ansatz}, we directly harvest data from atomistic simulations to statistically characterize the response of a sheared Cu$_{64}$Zr$_{36}$ glass.

\begin{figure}[t]
    \centering
    \includegraphics[width=8.0cm]{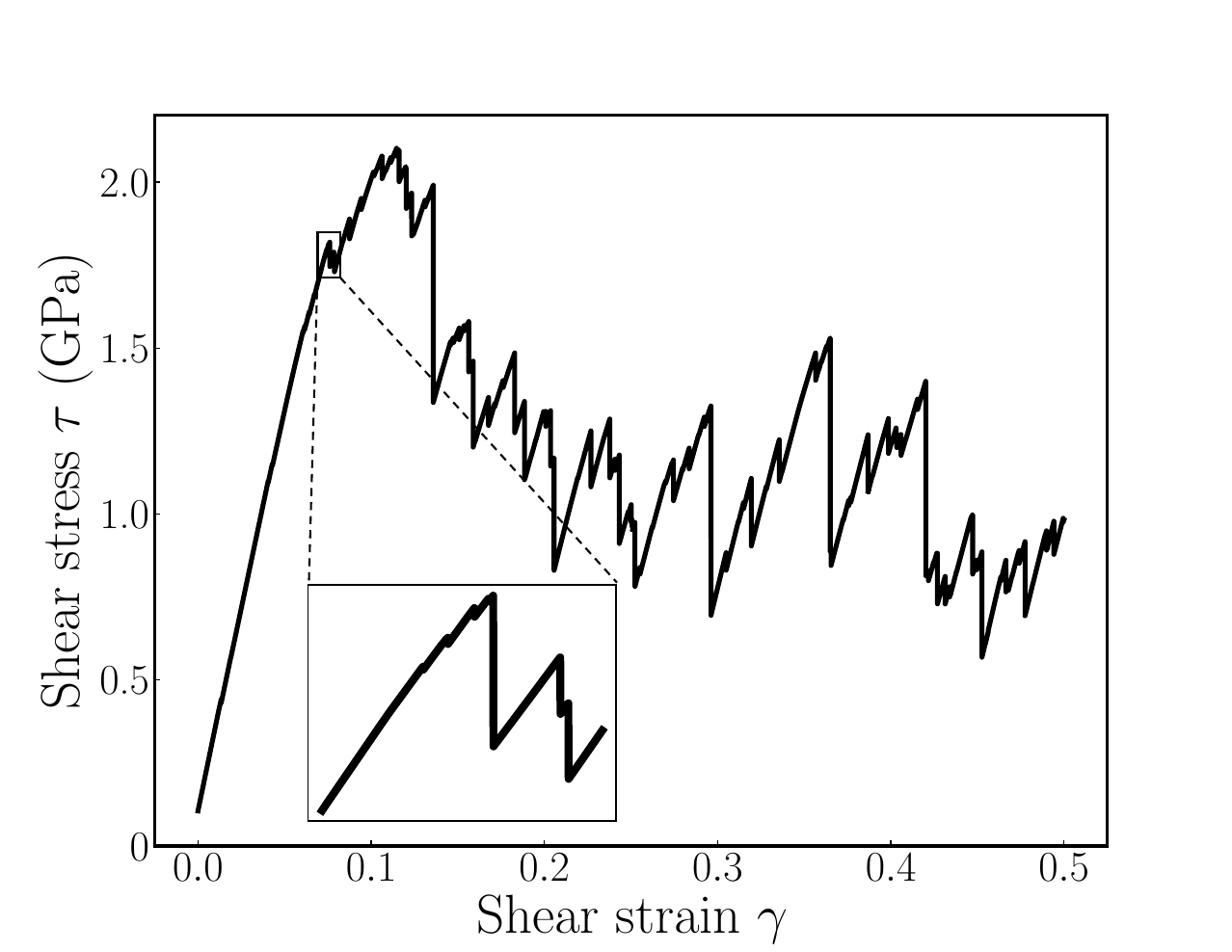}
    \includegraphics[width=8.0cm]{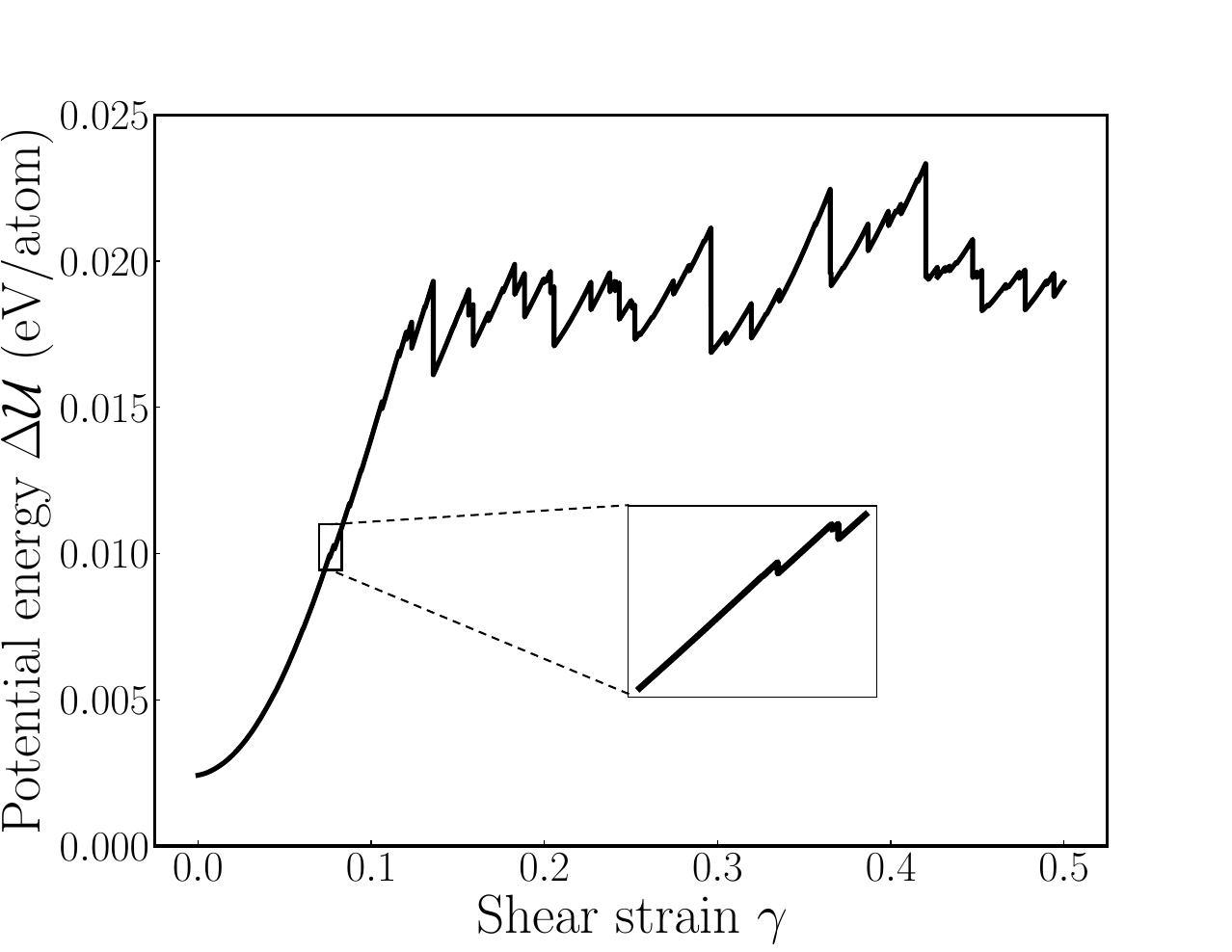}
    \caption{(a) Athermal quasistatic shear stress vs.\@ strain response of the simulated model of a Cu$_{64}$Zr$_{36}$ glass; (b) Athermal quasistatic potential energy vs.\@ strain response of this same model.}
    \label{fig:athermal_quasistatic_shear}
\end{figure}

A typical stress--strain curve obtained from an AQS protocol consists of a punctuated sequence of elastic branches, as shown in Fig.\,\ref{fig:athermal_quasistatic_shear}. 
 Increasing the strain along an elastic branch eventually induces a mechanical instability, noticeable in a discontinuous sudden jump in the potential energy $\mathcal{U}$ and a corresponding release in shear stress $\tau$ \cite{Bamer2023MolecularSolids}. Along any given elastic branch, the shear stress is defined by the differential relation $\tau(\gamma)=\frac{1}{V}\frac{d\mathcal{U}}{d\gamma}$, where $V$ denotes the volume of the simulation cell. Hence, the potential energy change during a strain increment $\Delta\gamma$ may be expressed
\begin{align}
    \label{eq:pot_energy_change}
    \Delta\mathcal{U} &= \mathcal{U}(\gamma+\Delta\gamma) - \mathcal{U}(\gamma) \nonumber\\  
    &= V\,\int_\gamma^{\gamma+\Delta\gamma} \tau(\gamma') d\gamma'+ 
    \Delta\mathcal{U}_\text{pl} \; \; .
\end{align}
If we assume that the shear modulus $G$ is constant during one strain step, the corresponding shear stress change is written
\begin{align}
    \label{eq:shear_stress_change}
    \Delta\tau &= \tau(\gamma+\Delta\gamma) - \tau(\gamma) \nonumber\\ 
    &= G\Delta\gamma  - \Delta\tau_\text{pl} \; \; .
\end{align}
The first terms on the right-hand side of Eqs.~\eqref{eq:pot_energy_change} \& \eqref{eq:shear_stress_change} describe the changes in potential energy and stress due to elastic work, and the second terms are related to the plastic rearrangements during the strain interval resulting in the release of elastic energy. In such a formulation, all thermodynamically irreversible processes are attributed to these plastic events during which the stress drops.

For the sake of generality, we rewrite ~\eqref{eq:pot_energy_change} \& \eqref{eq:shear_stress_change} in a matrix form in terms of the stress and the structural state of the system $\bm{X}$, where $\bm{X}= \{X_1,\dots, X_M\}$ captures all $M$ relevant degrees of freedom independent of stress. The shear modulus $G$ may also be dependent on the structural state $\bm{X}$.
This results in the expression
 \begin{align}
 \label{eq:raw-eq}
 \left( 
\begin{array}{c}
 \tau (\gamma+\Delta\gamma)-\tau(\gamma) \\
 X_1 (\gamma+\Delta\gamma)-X_1(\gamma) \\
 \vdots \\
 X_M (\gamma+\Delta\gamma)-X_M(\gamma)
\end{array}
 \right)
 = 
 \left( 
 \begin{array}{c}
G(\bm{X})\Delta\gamma \\
0 \\
\vdots \\
0 \\
 \end{array}
 \right) + \nonumber \\
 \sum_{i=1}^{n} 
 \left( 
\begin{array}{c}
 -\Delta\tau_{pl,i} \\
 \Delta X_{1,i}\\
 \vdots \\
 \Delta X_{M,i}
\end{array}
 \right) \; .
 \end{align}
The second term on the right-hand side sums over $n$, which denotes the number of plastic events during the strain interval.

Additionally, we assume that the number of plastic events $n$ during the strain interval $(\gamma,\gamma+\Delta\gamma)$ may be described by a Poisson distribution characterized by the state of the system, i.e.\@ $\tau$ and $\bm{X}$, so long as the interval is sufficiently small that this state does not appreciably vary \cite{Ahrens1974ComputerDistributions}.
Since the strain interval $\Delta\gamma$ can be chosen to be arbitrarily small, we will further simplify our numerical model by choosing an interval sufficiently small that the probability, $p=\lambda\Delta\gamma$, of an event in any interval may be assumed to be much less than unity, where $\lambda$ is a rate parameter describing the mean number of plastic events per unit strain. Equivalently, the probability of multiple plastic events in a single strain increment is vanishingly small. In this limit, the Poisson distribution of $n$ may be approximated by a Bernoulli distribution, and~\eqref{eq:raw-eq} can be simplified to
\begin{align}
 \label{eq:single}
 \left( 
\begin{array}{c}
 \tau (\gamma+\Delta\gamma)-\tau(\gamma) \\
 X_1 (\gamma+\Delta\gamma)-X_1(\gamma) \\
 \vdots \\
 X_M (\gamma+\Delta\gamma)-X_M(\gamma)
\end{array}
 \right)
 \approx 
 \left( 
 \begin{array}{c}
G(\bm{X})\Delta\gamma \\
0 \\
\vdots \\
0 \\
 \end{array}
 \right) + \nonumber \\
 \Theta\left[r-\lambda(\tau,\bm{X})\Delta\gamma\right]
 \left( 
\begin{array}{c}
 -\Delta\tau_{pl,\alpha} \\
 \Delta X_{1,\alpha}\\
 \vdots \\
 \Delta X_{M,\alpha}
\end{array}
 \right) \; .
 \end{align}
Here $\Theta$ is the Heaviside step function, and $r$ is a uniformly distributed random number on the interval from zero to one. The subscript $\alpha$ is used to label each plastic event, which is assumed to be unique.

The remaining challenge is to determine the changes in the stress and structural state of the system, i.e., the matrix on the far right side of~\eqref{eq:single}, and the rate of these changes $\lambda$. In the most general model, these depend on $\tau$ and $X_1,\dots, X_M$. While the literature on avalanche-like events in sheared disordered systems \cite{Kumar2022MappingModel, Salerno2013EffectDimensions} provides evidence that drops in both energy and stress during plastic events may be well-described by truncated power-law distributions, we will avoid making this assumption. Rather, we will directly harvest the statistics of plastic events by drawing from our database to inform our elasto-plastic modeling.

As a first step toward building a SEEM equivalent of our atomistic system, we begin by assuming that it is possible to characterize the state of the SEEM with only two state variables, $\tau$, and a single structural parameter $X_1$. The potential energy per atom, $\mathpzc{u}\equiv \mathcal{U}/N$, has been used in prior work to quantify important aspects of glass structure in atomistic simulations and has been related to the concept of the glass effective temperature \cite{Shi2007EvaluationSolids,Hinkle2017CoarseSolids,Alix-Williams2018ShearGlasses,Kontolati2021}. We choose $X_1$ to be the \textit{inelastic} potential energy per atom that we obtain by subtracting the elastic energy from the potential energy. This choice is made because the resulting quantity is physically meaningful and independent of the stress. The resulting expression for $X_1$ is
\begin{equation}  
    \label{eq:pe_in}
    X_1 = \mathpzc{u} - \mathpzc{u}_0 - \mathpzc{u}_\text{el}(\tau) \; ,
\end{equation}
where $\mathpzc{u}_0$ is an arbitrary reference energy chosen for convenience to be approximately the stress-free potential energy per atom of our most deeply quenched glass, and $\mathpzc{u}_\text{el}$ is the elastic energy per atom, a function of the shear stress $\tau$,
\begin{equation}  
    \label{eq:u_el_raw}
    \mathpzc{u}_\text{el} = \frac{\Omega\tau^2}{2 G_0} \left[ \frac{1}{G/G_0} + \left(\frac{\tau}{\tau_A}\right)^2+ \left(\frac{\tau}{\tau_B}\right)^4 \right] \; .
\end{equation}
Here, we approximate the elastic energy density with a truncated Taylor expansion, including only even terms to reflect the symmetry of the relationship between the elastic energy and the shear stress. $\Omega \equiv V/N $ is an inverse number density, i.e., the average volume per atom. We can further approximate this expression to reflect the fact that the shear modulus, $G$, depends on the structural state, $X_1$, 
\begin{equation}
    \label{eq:u_el}
    \mathpzc{u}_\text{el} \approx \frac{\Omega\tau^2}{2 G_0} \left[ \frac{1}{1 - X_1/\mathpzc{u}_A} + \left(\frac{\tau}{\tau_A}\right)^2+ \left(\frac{\tau}{\tau_B}\right)^4 \right] \; .
\end{equation}

The first term in the brackets of~\eqref{eq:u_el} arises from our empirical observation that the linear elastic modulus $G$ decreases with increasing $X_1$. This effect is quantified by extracting the parameters $G_0$ and $\mathpzc{u}_A$ from the stress--strain response for shear strains of less than $2\%$, as shown in Fig.~S1 in the supporting information. We obtain estimates of $\tau_A$ and $\tau_B$ by fitting them to the loading data from a large $400\,000$-atom AQS simulation to avoid the stochastic variations present in the smaller atomistic models. The complete details of the fitting are provided in Sec.~D in the supporting information, and the fitted values of the parameters are given in Table~\ref{table:parameter values}. The resulting relation between the structural state $X_1$ and the potential energy $\mathcal{U}$ and stress $\tau$ can be expressed in an approximate closed form as 
\begin{equation}
\label{eq:approximate closed form}
    X_1 \approx \frac{\mathpzc{u} - \mathpzc{u}_0 - \frac{\Omega\tau^2}{2 G_0} \left[1+ \left(\frac{\tau}{\tau_A}\right)^2+ \left(\frac{\tau}{\tau_B}\right)^4\right]}{1+\frac{\Omega \tau^2}{2G_0\mathpzc{u}_A}} \; .
\end{equation}

During the shear deformation protocol, every sample visits many states $(X_1,\tau)$.
Although this two-state variable model is an over-simplification, the process of generating and analyzing the results is interesting and informative. In doing so, we provide a baseline from which one could undertake iterative refinements to generate more adequate SEEMs that consider additional degrees of freedom in $\bm{X}$. 
These might, for example, be chosen based on physical intuition or by using machine learning techniques to optimally identify candidate degrees of freedom orthogonal to those in less refined iterations of the model.
\begin{table}[h!]
\centering
\begin{tabular}{ |p{2cm}||p{3cm}| }
 \hline
 Parameter& Fitted value\\
 \hline
 $\mathpzc{u}_A$ & $0.101\pm 0.002$ eV/atom\\
 $G_0$ & $25.90\pm 0.06$ GPa\\
 $\tau_A$ &  $8.4 \pm 0.5$ GPa  \\
 $\tau_B$& $2.90\pm 0.03$ GPa \\
 \hline
\end{tabular}
\caption{The parameters used to calculate $X_1$ and their fitted values.}
\label{table:parameter values}
\end{table}

\begin{figure*}[t]
    \centering
    \includegraphics[width=8.0cm]{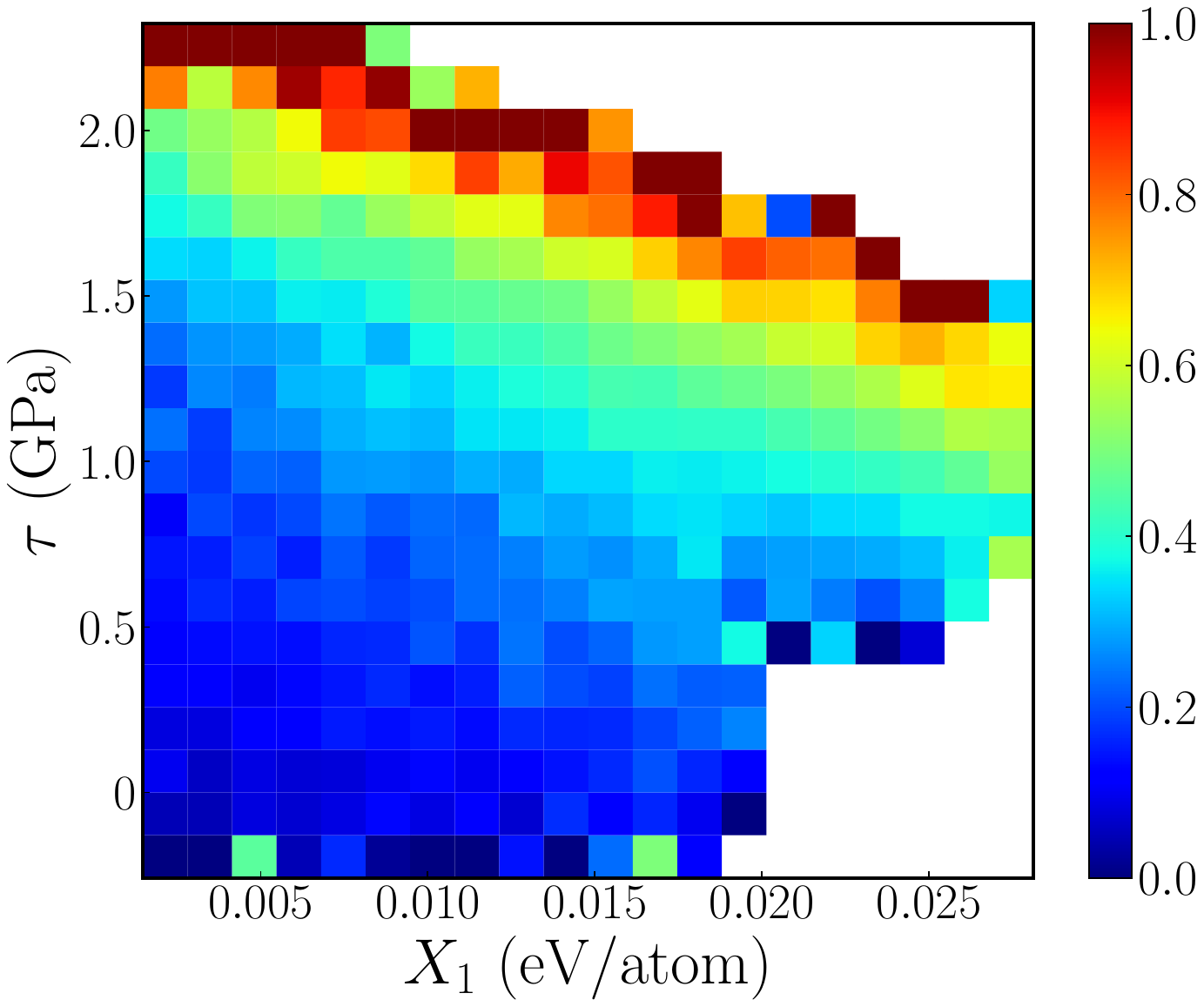}
    \includegraphics[width=8.0cm]{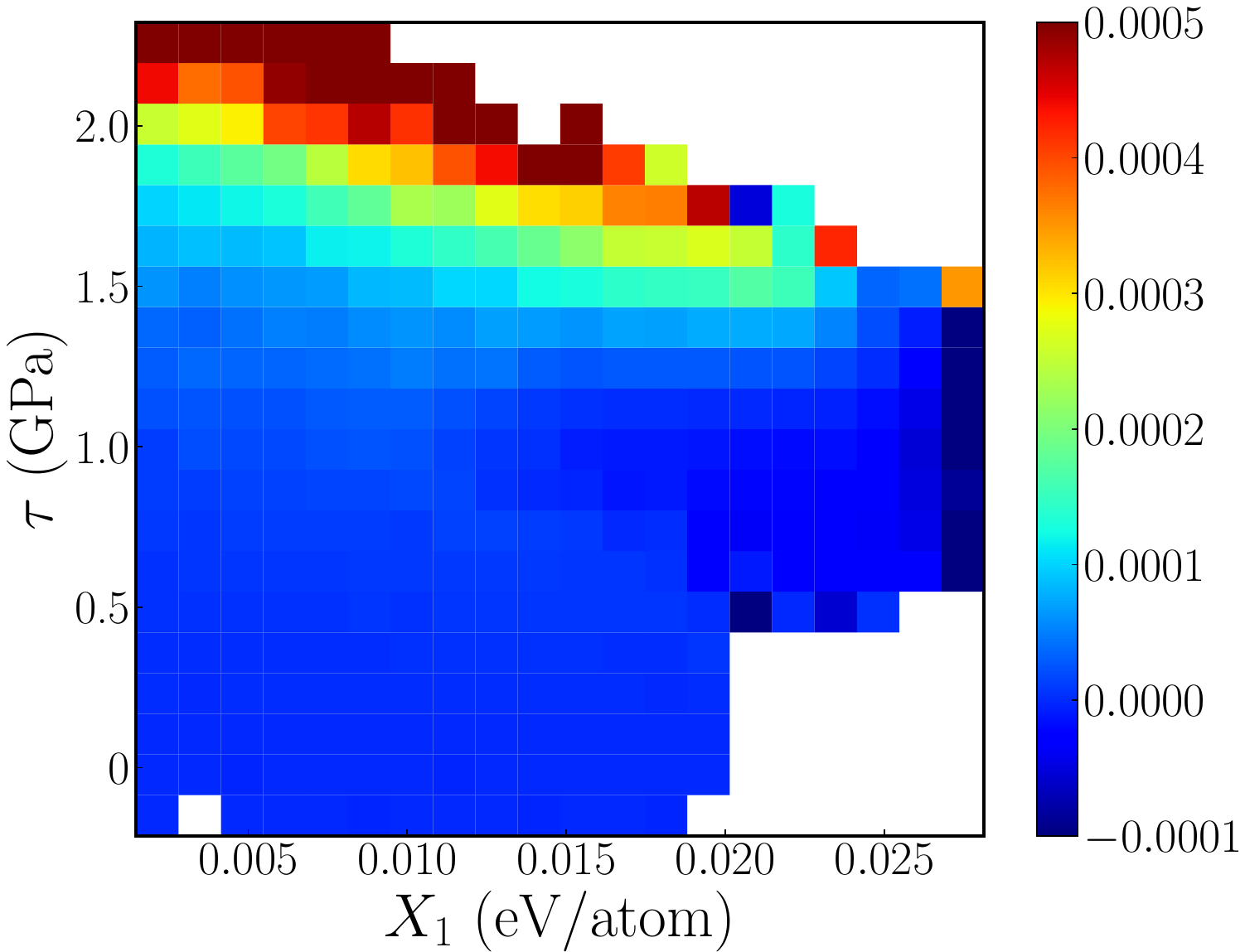}
    \includegraphics[width=8.0cm]{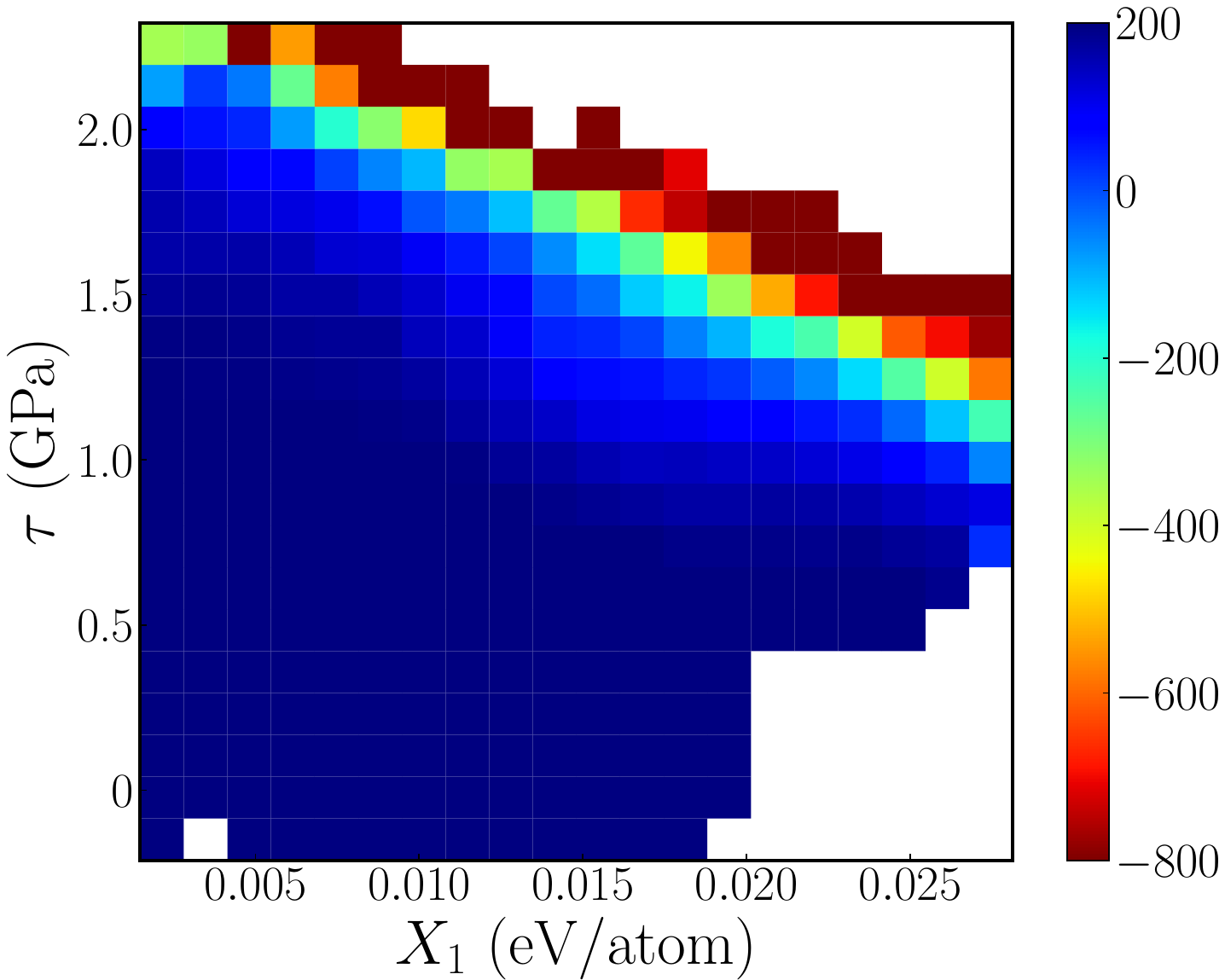}
    \includegraphics[width=8.0cm]{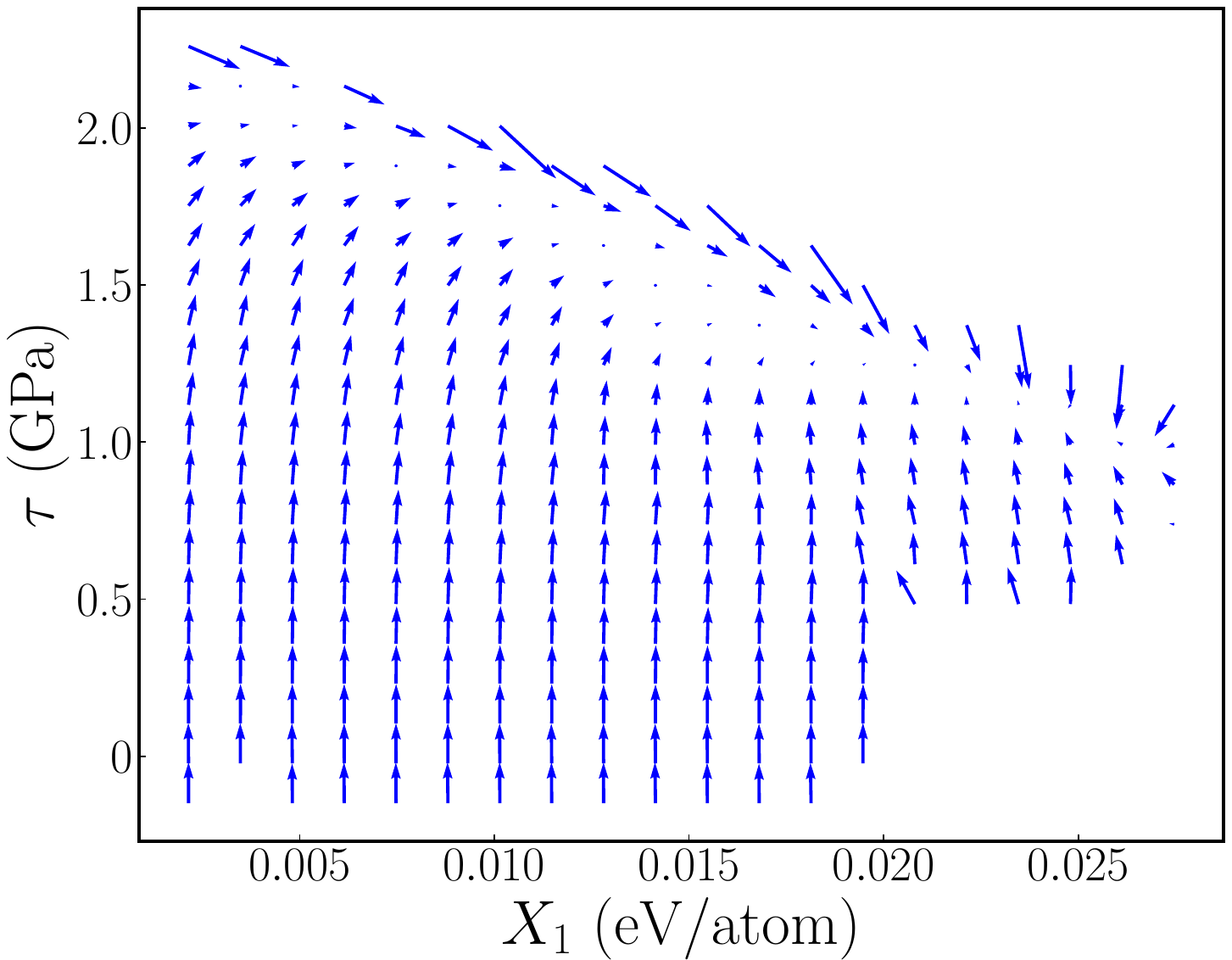}
    \caption{(a) Number of plastic events per strain increment divided by the total number of states harvested in every bin; (b) average inelastic potential energy change per plastic event; (c) average stress change per plastic event; (d) dynamic map of the gradient of inelastic potential energy change and stress change $(\frac{dX_1}{d\gamma},\frac{d\tau}{d\gamma})$.}
    \label{fig:states}
\end{figure*}

We generate a statistical database of the MG response behavior by harvesting stress drops and structural state changes from events across all visited state points.
For each visited state $(X_1,\tau)$, Fig.~\ref{fig:states} shows, per 0.1\% strain, (a) the average probability of a plastic event, (b) the average change in inelastic potential energy $\Delta X_1$ per plastic event
, and (c) the average change in the shear stress $\Delta\tau$ per plastic event.
The discretization of the state space is chosen to be $20\times 20$.  It is clear from the figures that all three of these quantities are highly correlated and that events become more frequent at high stress and large $X_1$, inducing an increase in the inelastic potential energy of the structure and triggering a net drop in stress. It is worth noting that some regions of the state space are never visited in our training data, particularly at high values of $X_1$ for both high and low values of stress. 

We note that $\Delta\tau$ and $\Delta X_1$ may both be either positive or negative. Obviously, $\Delta\tau$ is positive during elastic loading and negative during a plastic event. The change in the inelastic potential energy is positive during events that result in structurally less stable states, sometimes referred to as "rejuvenating". A negative change in the inelastic potential energy refers to aging events that result in structurally more stable states \cite{Lacks2004EnergyGlass}.
Figure \ref{fig:states}(d) shows the resulting dynamical map of the mean flow taking into account the change in the state per unit shear strain, i.e., $\left(\frac{dX_1}{d\gamma},\frac{d\tau}{d\gamma}\right)$. Such a map can be used to estimate the nullclines in $\Delta\tau$ and $\Delta X_1$ and to indicate the mean flow of the system through the state space during shear. The presence of an attractor is evident in the mean flow at the point where rejuvenation and aging events balance such that the net stress drop per unit strain precisely counters the elastic loading.   

\begin{figure*}[t]
    \centering
    \begin{tikzpicture}
        \node at (0,0){
            \includegraphics[width=10cm]{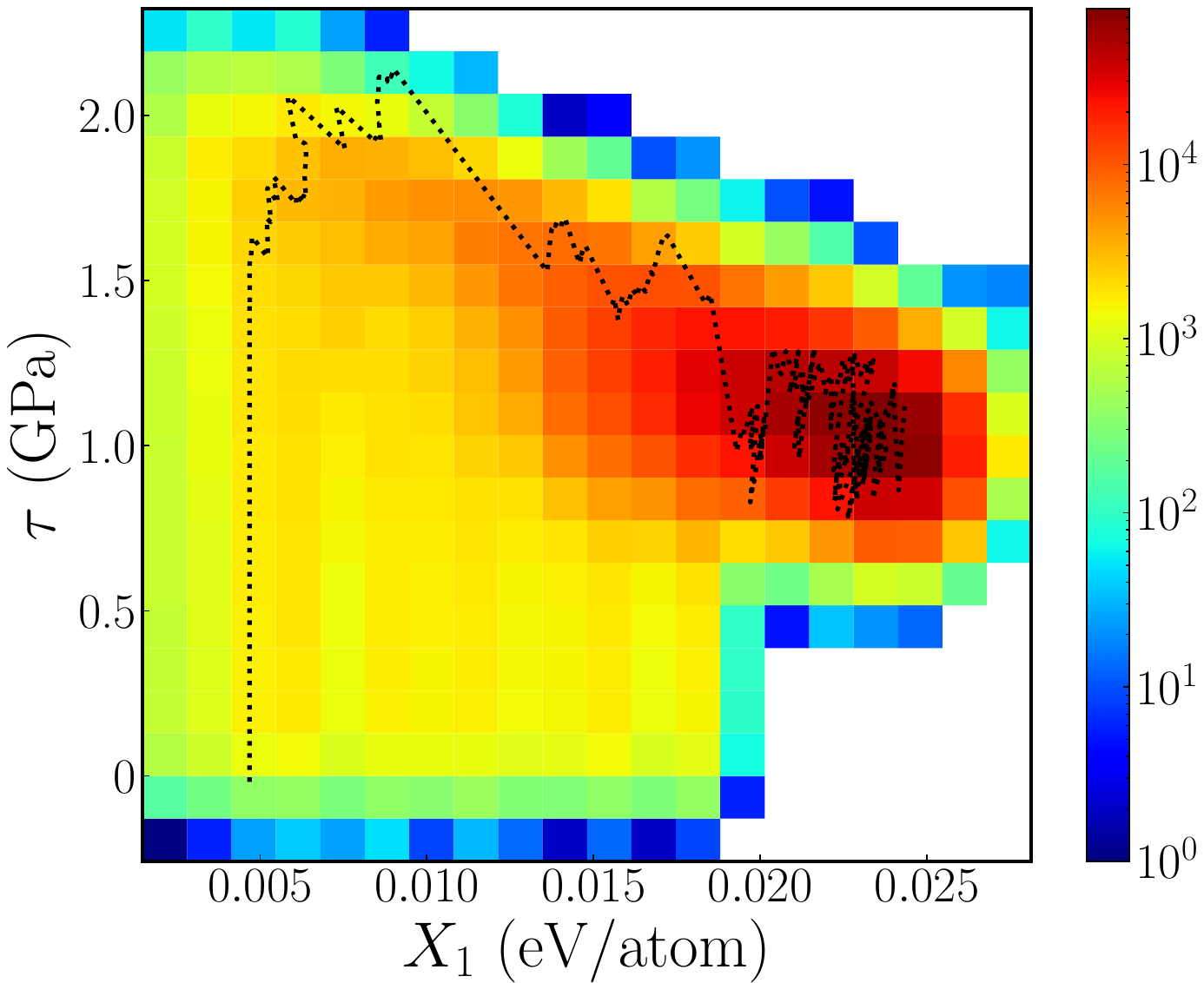}
        };
        \node[draw,color=black,line width=1pt, rounded corners, dotted] at (8,0){
            \includegraphics[width=6cm]{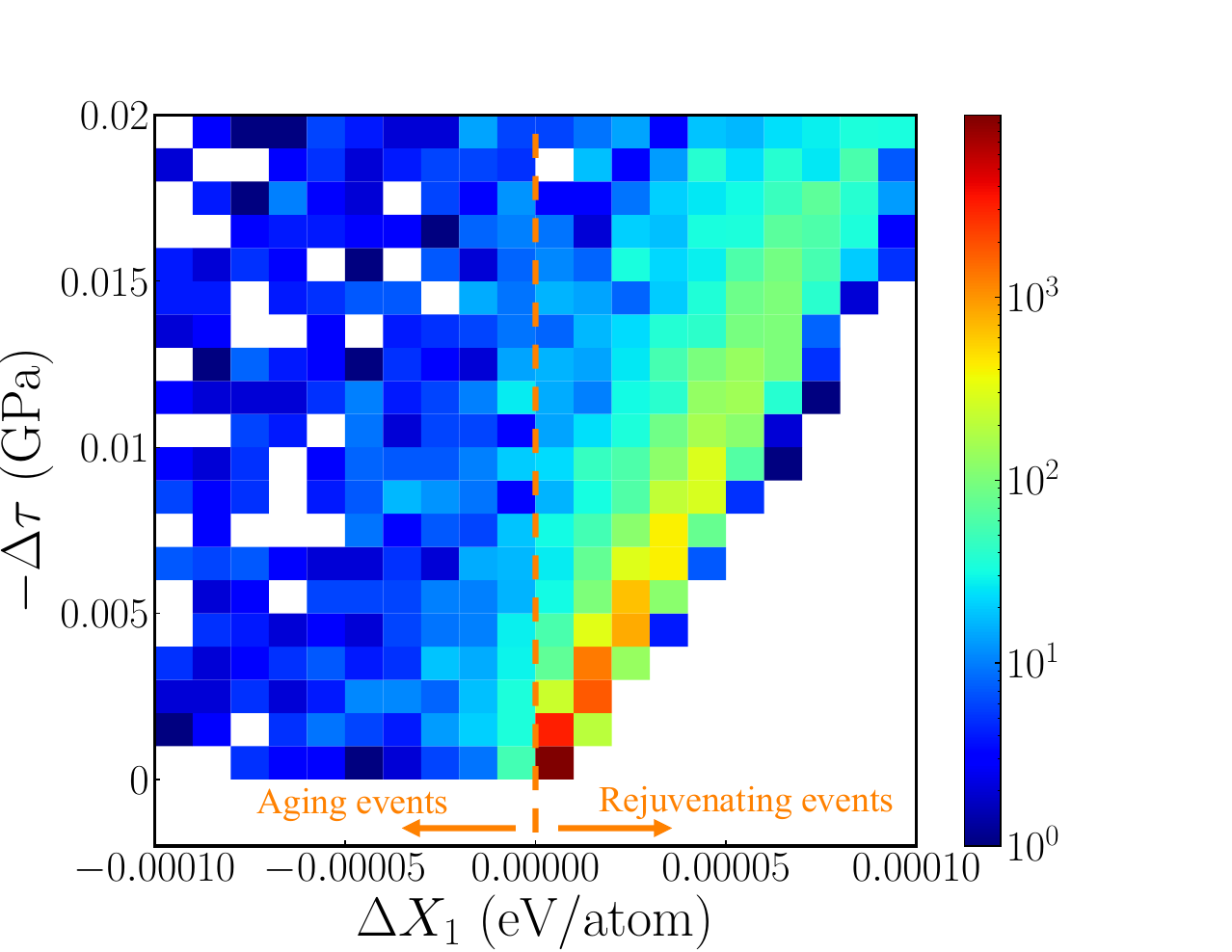}
        };
        \draw[-latex,color=magenta, line width=1.2pt] (1.6,0.121) --++ (3.2,0);
        \draw[color=magenta, line width=1.0pt] (1.2, -0.043) --++ (0.4,0) --++ (0,0.3) --++ (-0.4,0) -- cycle;
        \node at (-0.5,3.2){\textbf{(a)}};
        \node at (8,3.0){\textbf{(b)}};
    \end{tikzpicture}
    \caption{(a) Trajectory of an example of an $(X_1,\tau)$-evolution during deformation and total number of plastic events per strain at every state; (b) total number of plastic events depending on inelastic potential energy change $\Delta X_1$ and stress drop size $\Delta\tau$ in the most highly sampled state of the yielding region.}
    \label{fig:stress_drop_types}
\end{figure*}

The SEEM we propose is stochastic by construction. In the following, we describe the required data set to perform equation-free integration runs, i.e., the Markov chain of jumps from each state $(\tau^{(i)},X_1^{(i)})$ at shear strain $\gamma_i$ to a subsequent state $(\tau^{(i+1)},X_1^{(i+1)})$ at shear strain $\gamma_{i+1}=\gamma_i+\Delta\gamma$.
An essential feature of the SEEM is that the probability of an event at a particular state $(X_1,\tau)$ is determined by the simulation data collected. The average number of stress drops at particular states per 0.01\% strain is presented in Figure \ref{fig:states}(a). This determines the probability of a plastic event occurring during a strain increment, i.e., $\lambda\Delta\gamma$ in~\eqref{eq:single}. 
Figure \ref{fig:stress_drop_types}(a) shows the evolution of one shear deformation response in the $(X_1,\tau)$-space by the black dotted line superimposed on the state space colored by the number of stress drops analyzed in the statistical data set. 
The most populated state, which lies in the steady-state region, is represented in Figure \ref{fig:stress_drop_types}(b) by a two-dimensional histogram of stress drop events recorded within this region of state-space. Each such event contributes to the value of this histogram at its corresponding stress drop magnitude $\Delta\tau$ and change in inelastic potential energy $\Delta X_1$. The vertical dashed orange line denotes the $\Delta X_1=0$ axis and separates aging from rejuvenating events. While aging events are somewhat uniformly distributed over the negative region of $\Delta X_1$, the rejuvenating events are constrained to sit inside a linear boundary. This boundary describes the limit at which all elastic energy is converted into structural energy and, hence, no elastic energy is dissipated as heat. For events precisely on this boundary, the event would be ideally reversible, and the change in entropy $\Delta S=0$. 

Having described this single histogram (one of the approximately 400 histograms that comprise the database for our model), we now discuss how to perform the equation-free forward stepping from $(X_1^{(j)},\tau^{(j)})$ to $(X_1^{(j+1)},\tau^{(j+1)})$, which corresponds to a strain increment from $\gamma_j$ to $\gamma_{j+1}=\gamma_j+\Delta\gamma$. 
We start from an initial state $(X_1^{0},0)$. Using the value of $\lambda$ associated with this region of the state space, derived from the data in Fig.~\ref{fig:states}(a), we apply~\eqref{eq:single} to determine if a plastic event is triggered. 
If an event is not triggered, we evaluate the subsequent stress value $\tau^{(j+1)}=\tau^{(j)}+G(X_1^{j})\,\Delta\gamma$ using the relation between $G$ and $X_1$ derived in the supplementary information section (see Fig.~S1); the inelastic potential energy remains unaltered $X_1^{(j+1)}=X_1^{j}$. If an event is triggered, the histogram in the corresponding bin at state $(X_1^{(j)},\tau^{(j)})$ is sampled to select the stress drop magnitude and inelastic potential energy change during this strain interval. 
\begin{figure*}
    \centering
    \begin{tikzpicture}
        \node at (0,0){
            \includegraphics[width=8cm]{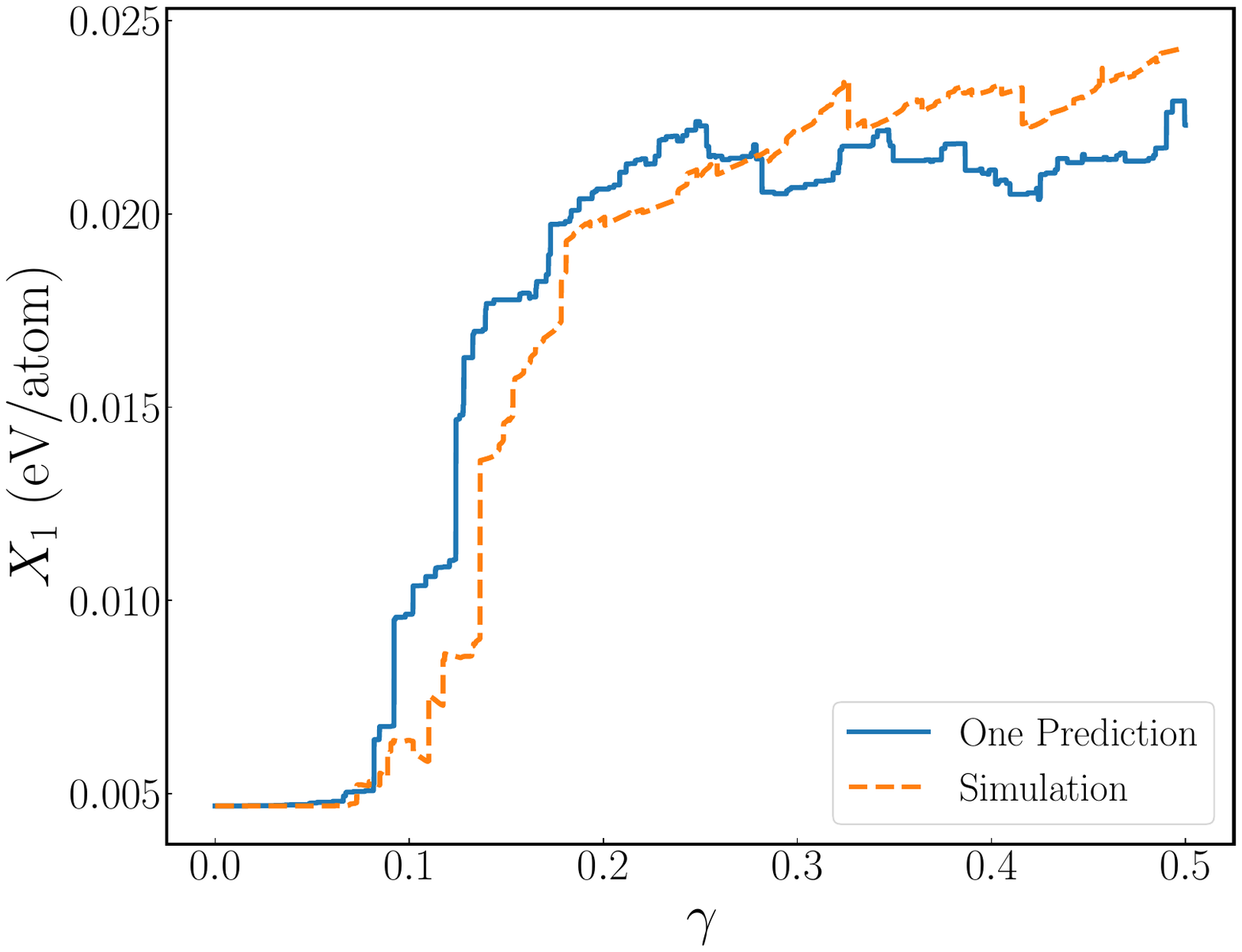}
        };
        \node at (8.2,0){
            \includegraphics[width=8cm]{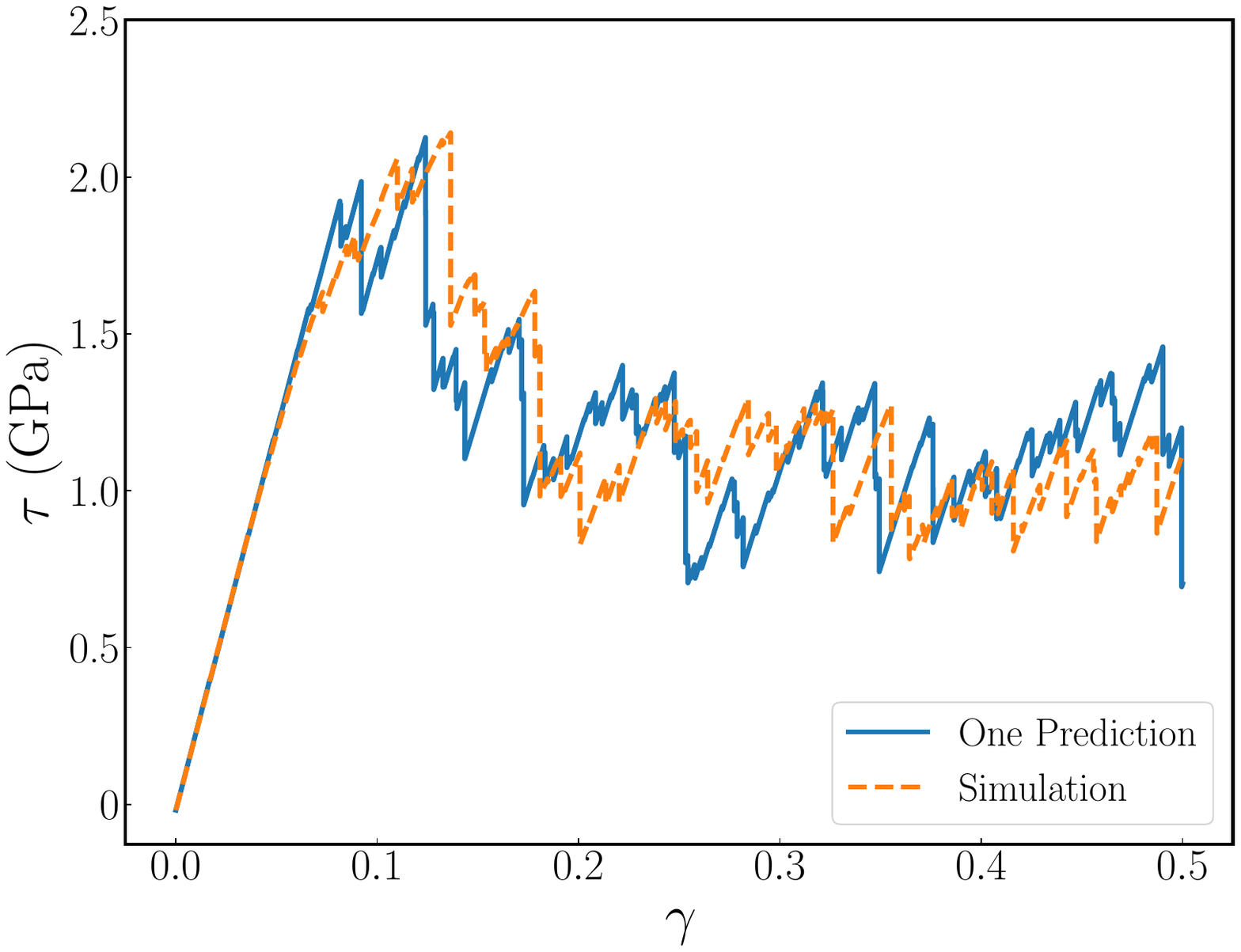}
        };
        \node at (0.32,3.2){\textbf{(a)}};
        \node at (8.2,3.2){\textbf{(b)}};
    \end{tikzpicture}
    \caption{Comparison of the inelastic potential energy results (a) and the stress--strain results (b) using the equation-free model (solid blue curve) and the atomistic simulation (dashed orange curve) starting from the same initial state. The $X_1$ values for the simulation are calculated using Eq.~\ref{eq:approximate closed form}.}
    \label{fig:stepping_res}
\end{figure*}

The black dotted line in Figure \ref{fig:stress_drop_types}(a) shows the result of one SEEM deformation run. After an initial stage of a nearly elastic response corresponding to a vertical trajectory, the system experiences jumps that increase the inelastic potential energy, i.e., rejuvenating events. This continues until the system approaches the steady-state attractor, at which point increases and decreases in inelastic potential energy, i.e., rejuvenating and aging events, balance. Generally, the system may start from any point in the $(X_1,\tau)$-space and will ultimately achieve a steady state stochastically exploring the region near this point in the state-space. 

Figure \ref{fig:stepping_res} compares the results from a SEEM run with the results using the conventional AQS simulation. Figure \ref{fig:stepping_res}(a) presents the evolution of the first state-variable, i.e., the inelastic potential energy, with respect to strain. Both results exhibit qualitatively similar behavior, an initial phase during which rejuvenation dominates and a second phase during which rejuvenating and aging balance.
Figure \ref{fig:stepping_res}(b) presents the evolution of the second state variable, i.e., the shear stress, with respect to shear strain. 
Our $4000$ atom system's behavior is captured surprisingly well by the SEEM with only two dynamical parameters.

\begin{figure}[h]
    \centering
    \includegraphics[width=10cm]{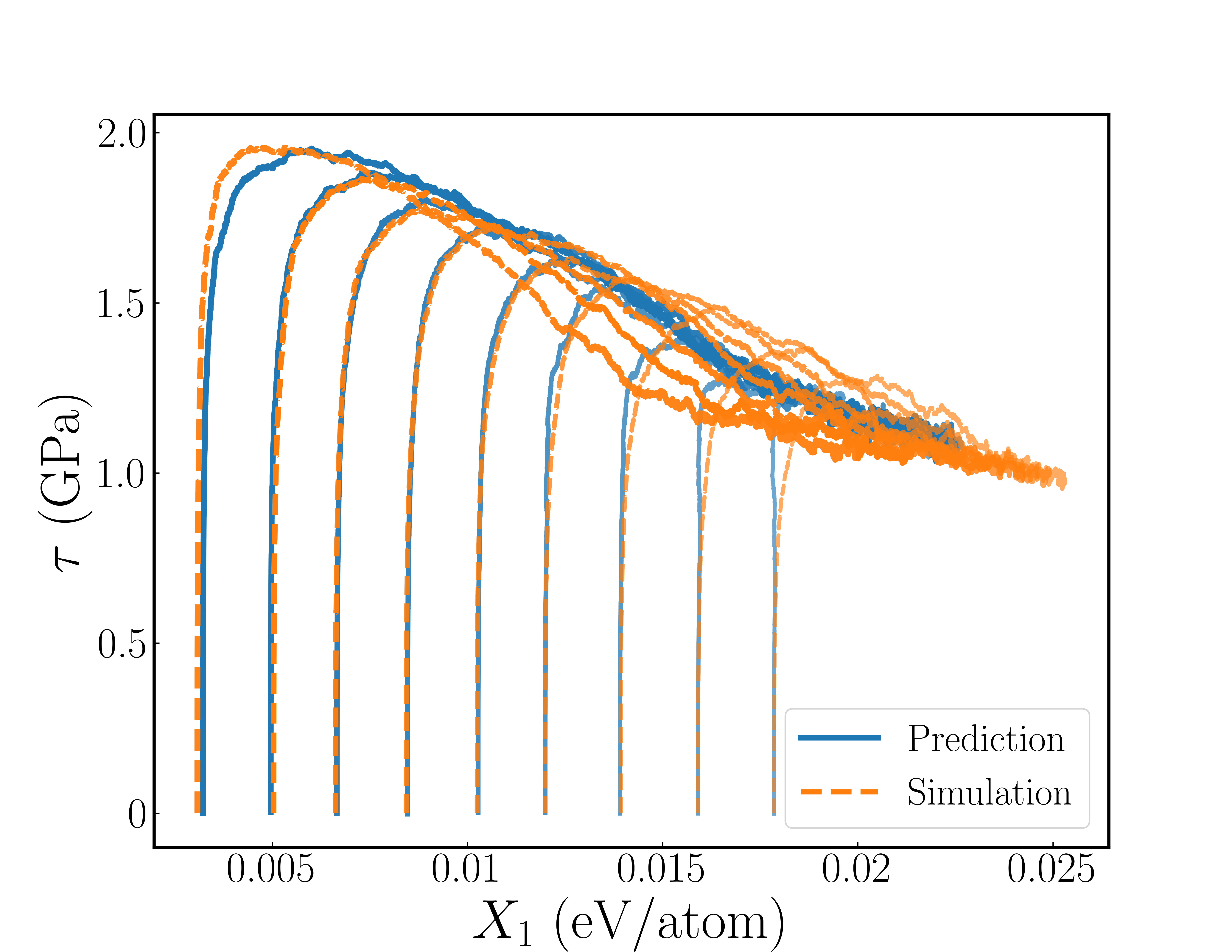}
    \label{fig:compare_one_simulation_stress}
    \caption{Comparison of the mean results of the average $(X_1,\tau)$ results applying different sets of cooling rates using the equation-free model (solid blue curve) and the classical molecular simulation (dashed orange curve). $X_1$ is calculated for the simulation using Eq.~\ref{eq:approximate closed form}.}
    \label{fig:average_res_cooling_rates}
\end{figure}

Nonetheless, some limits of the SEEM are evident in the resulting data. Figure \ref{fig:average_res_cooling_rates} presents the average state-space flow of the $(X_1,\tau)$-evolution of the material initiated from structures with different cooling rates. Lower cooling rates correspond to lower initial values of $X_1$. A crucial effect of the cooling rate on the $(X_1,\tau)$-evolution is that, with the increasing cooling rate, the stress peak value decreases. Notably, we see that the model prediction necessarily has the property that the trajectories in state-space do not cross one another. By construction, if two model trajectories intersect, they must converge since the location in the state-space fully determines the future behavior of the system. However, we see that the data directly derived from AQS simulations does not have this property. This provides direct evidence that this two-parameter state-space is insufficient to fully capture the behavior of this glassy system and that additional state variables are needed.

We conclude that SEEM presented in this paper represents many of the complex stochastic features of a disordered high-dimensional system with only two state variables represented by the stress and the inelastic potential energy. The model is purely data-driven and builds on statistics harvested from direct atomistic simulation. The results point to the possibility of reducing a complex high-dimensional system to a description in a small number of state variables. 
Clearly, a significant advantage of the SEEM is the vastly increased computational speed that makes this methodology attractive for incorporation into multiscale frameworks. 
At the same time, we see that due to the high degree of dimension reduction not all physical features of the MG system are captured. It will be interesting to rigorously define additional state variables orthogonal to $\tau$ and $X_1$ that allow for iteratively improved descriptions of MG systems.

\section*{Materials and Methods}
The mechanical system from which we harvest the data is a well-established binary MG-forming system. We prepare independent samples of a Cu$_{64}$Zr$_{36}$ glass by simulated quenching at nine different cooling rates. For the description of the atomic interaction, we used the embedded atom method \cite{Sheng2011HighlyMetals}. The simulation cell is cubic with a length of $39.88$\,\r{A} containing $4000$ atoms.  The quenching protocols were performed using a Nos\'e--Hoover thermostat \cite{Evans1985TheThermostat}. To prepare the samples, the temperature of the system was first held constant at $2500$\,K for $0.5$\,ns to reach equilibrium. It was then cooled to $1400$\,K at a rate of $10^{11}$\,K\,s$^{-1}$. During this cooling process, the system stays in equilibrium since the relaxation time is relatively short at this high temperature. To achieve uncorrelated samples, the velocities of the atoms were resampled at $1400$\,K following a Boltzmann distribution and further relaxed for $100$\,ps.
Finally, the samples were quenched to $100$\,K with cooling rates of $2^{x}\times 10^{10}$\,K\,s$^{-1}$, where $x = 1,2\dots,10$. Subsequently, a mechanically stable static structure, i.e., an inherent structure, was found by minimizing the potential energy until a force tolerance of $2\times 10^{-4}$\,eV/\r{A} was achieved.
Fifty samples were generated for each of the nine cooling rates. AQS was then imposed on each sample by successively relaxing the system to an inherent structure in the potential energy landscape using the conjugate gradient method after imposing shear strain increments of $10^{-5}$ \cite{Maloney2004SubextensiveFlow}. The system was sheared until a maximum strain value of $0.5$. To harvest the required data for our equation-free model, the potential energy and the stress evolution of the inherent structures at each strain state were recorded for further processing.



\section*{Author Contributions}
B.X. and M.L.F. conceptualized and developed the model;  B.X., Z.W., and J.L. undertook simulation, data analysis and testing; B.X. wrote the original draft. All authors engaged in discussions, model refinement, and editing.

\section*{Acknowledgement}
This work was supported by the U.S. National Science Foundation under Grant No.\@ \mbox{DMREF} 2323718/2323719/2323720 and was carried out at the Advanced Research Computing at Hopkins (ARCH) core facility which is supported by the NSF under Grant No.\@ OAC 1920103. Furthermore, Franz Bamer acknowledges the support of the German Research Foundation under Grant No. 523939420.


\bibliography{paper}

\end{document}